\documentclass[11pt, a4paper]{article}
\usepackage[a4paper, top=2.5cm, bottom=2.5cm, left=2.5cm, right=2.5cm]{geometry}
\usepackage[english]{babel}

\usepackage{noto}
\usepackage{graphicx}
\usepackage{amsmath}
\usepackage{natbib}
\usepackage{amssymb}
\usepackage{bm}
\usepackage{booktabs}
\usepackage{microtype}
\usepackage{setspace}
\usepackage[colorlinks=true, linkcolor=blue, citecolor=blue, urlcolor=blue]{hyperref}
\title{\textbf{A Two Stage Quasi-Likelihood Estimation Method for High Dimensional Generalized Structural Equation Models}}
\author{\textbf{Mohammad W. Hattab}\thanks{Correspondence: mwhattab@unm.edu}}

\date{} 
\begin{document}

\maketitle

\begin{abstract}
Estimating high dimensional Generalized Structural Equation Models presents severe computational challenges. Traditional simultaneous estimators frequently suffer from numerical instability and prohibitive computational costs. Moreover, there are no tractable algorithms for families such as Poisson, negative binomial, and gamma. To overcome these limitations, this article introduces a Two Stage Quasi-Likelihood Expectation-Maximization framework. The proposed method isolates the structural model from the measurement model. First, it approximates the conditional distribution of the latent variables given the observed indicators. Second, it employs marginal quasi-likelihood estimating equations to evaluate the structural parameters, deriving the necessary conditional moments either exactly or through Monte Carlo integration. This approach completely avoids the need to evaluate the full joint likelihood. Extensive simulations demonstrate that our method drastically reduces computational runtime, providing a numerically stable framework that minimizes the mean squared error and structural bias to yield a scalable and flexible solution for analyzing complex latent variable models.

\vspace{0.5cm}
  
\end{abstract}
\doublespacing

\section{Introduction}

In many applications across the social, behavioral, and health sciences, researchers aim to model the relationship between a set of unobserved latent variables and an observed response variable. A standard framework for such analysis is the latent variable generalized linear model (LVGLM), also recognized as a Generalized Structural Equation Model (GSEM) or a Generalized Linear Latent Variable Model (GLLVM). Assessments in these fields often require the modeling of a high dimensional latent space where indicators consist of a heterogeneous mix of binary, ordinal, and continuous scales.

Consider a sample of $n$ independent observations. For each subject $i$ ($i = 1, \dots, n$), let $\bm{\eta}_i$ be a $p \times 1$ vector of unobserved latent predictors.  We assume that the latent predictors are sampled from a multivariate normal distribution:
\begin{equation}
\bm{\eta}_i \sim N(\bm{0}, \bm{\Phi}), \label{eq:eta_dist}
\end{equation}
where $\bm{\Phi}$ is a $p \times p$ symmetric, positive definite covariance matrix. These unobserved latent predictors are measured via a $J \times 1$ vector of observed indicators, $\bm{Z}_i = (Z_{i1}, \dots, Z_{iJ})^\top$. The measurement model relating the latent vector $\bm{\eta}_i$ to these indicators is typically defined as:
\begin{equation}
\bm{Z}_i = \bm{\nu} + \bm{\Lambda} \bm{\eta}_i + \bm{\epsilon}_i, \label{eq:measurement}
\end{equation}
where $\bm{\Lambda}$ is a $J \times p$ matrix of factor loadings and $\bm{\epsilon}_i$ is a $J \times 1$ vector of independent measurement errors with $\bm{\epsilon}_i \sim N(\bm{0}, \bm{\Psi})$. For categorical or ordinal indicators, this relationship is further defined via a set of thresholds mapping the underlying continuous scale to the observed discrete categories.

Let $y_i$ be the observed response variable. We formulate the structural relationship between the latent variables and the outcome as a Generalized Linear Model (GLM). In this framework, $y_i$ follows a distribution within the exponential family (or a quasi-likelihood family) with the structural equation given by:
\begin{equation}
g(\mu_i) = \beta_0 + \bm{\eta}_i^\top \bm{\beta}, \label{eq:structural}
\end{equation}
where $\mu_i = E(y_i \mid \bm{\eta}_i)$ is the conditional mean of $y_i$ given $\bm{\eta}_i$, $g(\cdot)$ is a known, monotonic, and differentiable link function, $\beta_0$ is the intercept, and $\bm{\beta}$ is a $p \times 1$ vector of unknown structural regression parameters.

The GSEM framework provides a unified system that encompasses item response theory, factor analysis, and generalized linear modeling (Breslow \& Clayton, 1993; Muthén, 2002; Skrondal \& Rabe-Hesketh, 2004). In clinical psychology, GSEMs are utilized to predict behavioral outcomes or binary diagnoses from highly correlated latent personality traits (Bartholomew et al., 2011). In epidemiology and public health, researchers employ these models to assess the impact of unobserved frailty indices measured via mixed categorical and continuous survey items on disease incidence and mortality rates (Sammel et al., 1997; Rabe-Hesketh et al., 2004). Furthermore, in educational testing, GSEMs allow for the prediction of drop-out probabilities or bounded test scores using multiple cognitive traits (De Boeck \& Wilson, 2004).

However, the estimation of high dimensional GSEMs presents a significant computational challenge. The observed data likelihood function involves an integral over the $p$-dimensional latent space:
\begin{equation}
L(\bm{\beta}, \bm{\Lambda} \mid \bm{Y}, \bm{Z}) = \prod_{i=1}^n \int f(y_i \mid \bm{\eta}_i, \bm{\beta}) f(\bm{Z}_i \mid \bm{\eta}_i, \bm{\Lambda}) f(\bm{\eta}_i) d\bm{\eta}_i. \label{eq:likelihood}
\end{equation}

When the model involves non-Gaussian outcomes, non-linear link functions, or categorical indicators, this integral does not result in a closed form analytical solution. While the integral can be approximated to high precision using standard numerical techniques---such as Gauss-Hermite quadrature---these methods scale exponentially with the number of latent dimensions $\mathcal{O}(k^p)$. Consequently, it becomes computationally prohibitive for high dimensional latent spaces. While stochastic approximations like the Metropolis--Hastings Robbins--Monro (MHRM) algorithm (Cai, 2010a, 2010b) have been proposed to resolve this, they remain computationally intensive and can become unstable as the dimensionality of the latent space and the complexity of the indicators increase. 

Simultaneous full information estimators, such as Full Information Maximum Likelihood (FIML; Arbuckle, 1996) and Bayesian MCMC (Arminger \& Muthén, 1998), jointly estimate all parameters to maximize asymptotic efficiency.  Because the entire system is evaluated jointly, these methods are highly vulnerable to interpretational confounding. A localized misspecification in the measurement model (Eq. \ref{eq:measurement}) may lead to severe bias in the structural parameters  (Eq. \ref{eq:structural} ) as the optimizer attempts to maximize global fit (see Anderson \& Gerbing, 1988; Carroll et al., 2006 ).

Additionally, FIML is highly efficient for continuous indicators but becomes mathematically intractable with discrete variables, as it requires evaluating complex multidimensional normal integrals. Bayesian approaches successfully avoid the need for numerical integration via stochastic sampling, however, they remain computationally exhaustive, requiring long Markov chains that often struggle to mix and converge as the parameter space expands.  Finally, simultaneous limited-information estimators, such as Robust Weighted Least Squares (WLSMV; Muthén, 1984), avoid the integration problem entirely by estimating the model from univariate and bivariate margins (e.g., polychoric correlations). However, they require the calculation of massive asymptotic weight matrices and threshold vectors, frequently triggering severe numerical instability and non-convergence in high dimensional categorical settings.

To overcome these computational limitations, this article introduces the Two-Stage Quasi-Likelihood Expectation Maximization (TSQLEM) framework. Rather than jointly estimating the entire system, the proposed method isolates the measurement model from the structural model. The first stage yields the conditional distribution of the latent space given the observed indicators, $\bm{\eta} \mid \bm{Z}$.  Following, it targets the conditional moments of the outcome given the observed indicators, $y \mid \bm{Z}$. By evaluating the conditional first and second moments via exact analytical derivations or Monte Carlo integration, the structural parameters are subsequently estimated using an Iteratively Reweighted Least Squares (IRLS) scheme. This framework provides a highly scalable estimator. Since it relies on quasi-score equations, the framework is flexible to accommodate any distribution and link function regularly considered in generalized linear models, including quasi-likelihood families.

The remainder of this article is organized as follows. Section 2 introduces the TSQLEM framework, detailing the derivation of the conditional moments, the construction of the approximate quasi-score equations, and the mechanics of the iterative estimation algorithm. Section 3 presents a comprehensive Monte Carlo simulation study designed to compare the proposed method against traditional full and limited information estimators. In Section 4, the framework is applied to an empirical dataset (the Big Five Inventory) to demonstrate its practical use in estimating a high dimensional structural model from ordinal indicators. Finally, Section 5 summarizes the findings.

\section{The Proposed Method: A Two stage Quasi-Likelihood EM  method}

Our objective is to derive or find an approximation of the conditional moments of $\bm{Y}$ given the observed indicators $\bm{Z}$.

Our framework shares the theoretical foundation of Marginal Maximum Likelihood (MML; Bock \& Aitkin, 1981) by explicitly integrating out the unobserved latent space. However, rather than evaluating the intractable joint likelihood to capture the full conditional distribution, we target only those necessary conditional moments and modularize the integral into a two stage approach.

To execute this, we propose a Two Stage Quasi-Likelihood Expectation-Maximization (TSQLEM) approach. This method operates by first isolating the measurement model to approximate the conditional distribution of the latent variables given the observed indicators. Subsequently, the structural model employs a marginal quasi-likelihood estimating equation relying on the derived conditional moments of the observed data.

For some models (e.g., binary probit), these conditional moments are derived via analytical integration. When those quantities cannot be analytically obtained, the TSQLEM framework approximates the conditional moments using Monte Carlo integration to yield an approximated marginal quasi-likelihood estimating equation. Furthermore, in order to obtain unbiased estimates of the quasi-likelihood weights, a Taylor expansion is applied utilizing the conditional variance to correct for simulation noise.

\subsection{Stage 1: The Measurement Model}

In the first stage, we fit a measurement model (e.g., Confirmatory Factor Analysis) to the observed indicators $\mathbf{Z}$. Assuming a standard linear measurement model (Eq. \ref{eq:measurement}), $\mathbf{Z}_i = \bm{\nu} + \bm{\Lambda} \bm{\eta}_i + \bm{\epsilon}_i$, we utilize Empirical Bayes to estimate the conditional distribution of the latent variables given the observed data, extracting $\hat{\bm{\eta}}_i$ and $\bm{\Sigma}_{i}$ as its conditional mean and covariance matrix, respectively.

 Depending on the nature of the indicators, this stage could  utilize various estimators. However, we recommend utilizing a continuous Maximum Likelihood (ML) approximation, even when indicators are categorical or ordinal. Under this estimation, $\bm{\Sigma}_{i}$ simplifies to a covariance matrix $\bm{\Sigma}$ that remains constant across all individuals. This simplification provides computational speed, which is invaluable when considering high dimensional models. While treating categorical indicators as continuous introduces a degree of misspecification at the measurement level, we demonstrate in the subsequent simulation study (Section 3) that this simplification is remarkably robust. This estimation framework approximates the distribution of $\bm{\eta}_{i}\mid \bm{Z}_i$ as $N(\hat{\bm{\eta}}_i, \bm{\Sigma})$.
 
Utilizing Monte Carlo integration in Stage 2 allows the TSQLEM framework to accommodate non-normal conditional distributions $f(\bm{\eta}_i \mid \bm{Z}_i)$. Should the researcher suspect severe skewness, multimodality, or heavy tails in the latent space, Stage 1 can utilize non-parametric alternatives. For instance, Bayesian Nonparametric Confirmatory Factor Analysis (BNP-CFA) with Dirichlet Process priors, or frequentist Factor Mixture Modeling (FMM), can be employed to flexibly capture the true  features of the latent space. The resulting conditional distributions by these non-parametric methods can be integrated directly into the TSQLEM estimating equations without altering the core of the Stage 2 algorithm.

\subsection{The Conditional Moments of the Response Variable}

 We cannot simply regress $y_i$ on $\hat{\bm{\eta}}_i$ without inducing severe attenuation bias. Instead, we must integrate over the uncertainty in the latent space by deriving the conditional expectation and variance of $y_i$ given the observed indicators $\bm{Z}_i$.

Due to the conditional independence of $y_i$ and $\bm{Z}_i $ given $ \bm{\eta}_i$, the conditional mean of $y_i \mid \bm{Z}_i$ is obtained via nested expectation:
\begin{equation}
E(y_i \mid \bm{Z}_i) = E\left[ E(y_i \mid \bm{Z}_i, \bm{\eta}_i) \mid \bm{Z}_i \right] = E\left[ g^{-1}(\beta_0 + \bm{\eta}_i^\top \bm{\beta}) \mid \bm{Z}_i \right]. \label{eq:cond_mean}
\end{equation}
Next, the conditional variance of $y_i \mid \bm{Z}_i$ is derived using the law of total variance:
\begin{align}
var(y_i \mid \bm{Z}_i) &= E\left[ var(y_i \mid \bm{Z}_i, \bm{\eta}_i) \mid \bm{Z}_i \right] + var\left( E(y_i \mid \bm{Z}_i, \bm{\eta}_i) \mid \bm{Z}_i \right) \nonumber \\
&= E\left[ V(\beta_0 + \bm{\eta}_i^\top \bm{\beta}, \phi) \mid \bm{Z}_i \right] + var\left( g^{-1}(\beta_0 + \bm{\eta}_i^\top \bm{\beta}) \mid \bm{Z}_i \right), \label{eq:cond_var}
\end{align}
where $V(\cdot, \phi)$ is the variance function specified by the assumed GLM family and $\phi$ is the dispersion parameter.

To evaluate these integrals, the TSQLEM engine employs a strategy depending on the nature of the link function. 

\subsubsection{General Link Functions: Monte Carlo Integration}
These quantities cannot be found analytically for most link functions. However, they can be accurately approximated via Monte Carlo integration. For each subject $i$, we generate $B$ random draws, $\bm{\eta}_{is}$ ($s = 1, \dots, B$), from the  multivariate normal conditional distribution:
\begin{equation}
\bm{\eta}_{is}\mid \bm{Z}_i \sim N(\hat{\bm{\eta}}_i, \bm{\Sigma}_{i}). \label{eq:mc_draws}
\end{equation}
Let $\bm{X}_{is} = (1, \bm{\eta}_{is}^\top)^\top$ be the design vector for draw $s$, and let $\bm{\gamma} = (\beta_0, \bm{\beta}^\top)^\top$ be the combined structural parameter vector. We approximate the conditional mean in Eq. \ref{eq:cond_mean} as:
\begin{equation}
\mu_{i} = E(y_i \mid \bm{Z}_i) \approx \frac{1}{B} \sum_{s=1}^B g^{-1}(\bm{X}_{is}^\top \bm{\gamma}). \label{eq:approx_mean}
\end{equation}
Similarly, the conditional variance in Eq. \ref{eq:cond_var} is approximated as:
\begin{equation}
V_{i} = var(y_i \mid \bm{Z}_i) \approx \frac{1}{B} \sum_{s=1}^B V(\bm{X}_{is}^\top \bm{\gamma}, \phi) + \widehat{var}\left( g^{-1}(\bm{X}_{is}^\top \bm{\gamma}) \right), \label{eq:approx_var}
\end{equation}
where the second term represents the sample variance computed across the $B$ evaluations of the inverse link function for subject $i$.

As mentioned earlier, the Monte Carlo integration does not require the latent space to follow a multivariate normal distribution, but rather it only requires a machinery to generate simulated draws that reasonably approximate the conditional distribution of the latent space.

\subsubsection{The Log and Probit Links: Exact Analytical Integration}
Log link functions are customarily used for Poisson, quasi-Poisson, Gamma, and Negative Binomial regressions. For those models  Monte Carlo integration is unnecessary. Instead, we derive exact and closed-form solutions via the Moment Generating Function (MGF) of the Multivariate Normal distribution.

Let $\bm{X}_{obs, i} = (1, \hat{\bm{\eta}}_i^\top)^\top$ be the observed design vector based on the expected latent score. To map the conditional covariance to the full parameter vector, we define an augmented covariance matrix $\tilde{\bm{\Sigma}}_i = \begin{bmatrix} 0 & \bm{0}^\top \\ \bm{0} & \bm{\Sigma}_i \end{bmatrix}$. It follows that the conditional mean and variance of the unobserved linear predictor $\beta_0 + \bm{\eta}_i^\top \bm{\beta}$ are $ \bm{X}_{obs,i}^\top \bm{\gamma}$ and $\sigma_{\eta, i}^2 = \bm{\gamma}^\top \tilde{\bm{\Sigma}}_i \bm{\gamma}$, respectively. 

Applying the MGF, the exact conditional mean from Eq. \ref{eq:cond_mean} resolves analytically to:
\begin{equation}
\mu_{i} = \exp\left( \bm{X}_{obs, i}^\top \bm{\gamma} + \frac{1}{2} \sigma_{\eta, i}^2 \right). \label{eq:exact_mean}
\end{equation}
Furthermore, the exact conditional variance is derived from Eq. \ref{eq:cond_var}:
\begin{equation}
V_{i} =  E\left[ V(\beta_0 + \bm{\eta}_i^\top \bm{\beta}, \phi) \mid \bm{Z}_i \right] + \mu_{i}^2 \left( \exp(\sigma_{\eta, i}^2) - 1 \right). \label{eq:exact_var}
\end{equation}

The first term in Eq. \ref{eq:exact_var} is simply $\mu_i$ under the Poisson family. The derivation for the Gamma family is given in Section 2.5.

Bernoulli models employing a probit link function also admit an exact analytical solution. Under the probit link,  $E(y_i\mid \bm{\eta}_i) =  \Phi(\beta_0 + \bm{\eta}_i^\top \bm{\beta})$, where  $\Phi(\cdot)$ is the cumulative distribution function of the standard normal distribution.

Because the convolution of a standard normal cumulative distribution function with a normal probability density function yields another normal CDF (see Murphy, 2012; Rasmussen \& Williams, 2006), the conditional mean is given via Eq. \ref{eq:cond_mean}, 

\begin{equation}
\mu_i = E \left[ \Phi(\beta_0 + \bm{\eta}_i^\top \bm{\beta}) \mid \bm{Z}_i \right] = \Phi\left(\frac{\bm{X}_{obs,i}^\top \bm{\gamma}}{\sqrt{1 + \sigma_{\eta,i}^2}}\right). \label{eq:mean_probit}
\end{equation}

For binary outcomes, the distribution of $y_i \mid \bm{Z}_i$ is Bernoulli as well. This immediately implies that 
\begin{equation}
V_i = var(y_i \mid \bm{Z}_i) =\mu_i (1-\mu_i).  \label{eq:varBern}
\end{equation}

Notice that there is no closed form solution for $V_i$ under Binomial models with $m > 1$ trials and one needs to resort to the Monte Carlo integrations described previously. More details are given in Section 2.4.


\subsection{Stage 2: The Generalized Iterative Scheme}

The estimation of the structural parameters relies on the framework of marginal quasi-likelihood. If the exact marginal mean $\mu_{i}(\bm{\gamma}) = E(y_i \mid \bm{Z}_i)$ and marginal variance $V_{i} = var(y_i \mid \bm{Z}_i)$ were analytically known (as is the case for the log-link with normally distributed latent predictors), we could linearize the exact mean via a first-order Taylor series expansion around the current estimate $\bm{\gamma}^{(t)}$:
\begin{equation}
    \mu_{i}(\bm{\gamma}) \approx \mu_{i}(\bm{\gamma}^{(t)}) + \bm{D}_{i}^{(t)\top} (\bm{\gamma} - \bm{\gamma}^{(t)}), \label{eq:exact_linearization}
\end{equation}
where $\bm{D}_{i}^{(t)} = \frac{\partial \mu_{i}(\bm{\gamma})}{\partial \bm{\gamma}}\Big|_{\bm{\gamma}=\bm{\gamma}^{(t)}}$ is the gradient vector. Substituting this expansion into the quasi-likelihood framework yields the exact marginal quasi-score equation:
\begin{equation}
    \bm{U}(\bm{\gamma}) = \sum_{i=1}^n \bm{D}_{i}^{\top} V_{i}^{-1} (y_i - \mu_{i}) = \bm{0}. \label{eq:exact_score}
\end{equation}

However, these exact conditional moments are analytically unattainable for most general link functions. Therefore, we must instead apply the Taylor series expansion directly to the Monte Carlo approximation of the mean. For every draw $s = 1, \dots, B$, let $\mu_{is}(\bm{\gamma}) = g^{-1}( \bm{X}_{is}^\top \bm{\gamma})$, $\mu_{is}^{(t)} = g^{-1}( \bm{X}_{is}^\top \bm{\gamma}^{(t)})$, and $\bar{\mu}_i(\bm{\gamma}) = \frac{1}{B} \sum_{s=1}^B \mu_{is}(\bm{\gamma})$. Expanding this simulated mean around $\bm{\gamma}^{(t)}$ yields:
\begin{equation}
    \bar{\mu}_i(\bm{\gamma}) \approx \frac{1}{B} \sum_{s=1}^B \mu_{is}^{(t)} + \left( \frac{1}{B} \sum_{s=1}^B \frac{\partial  \mu_{is}(\bm{\gamma})}{\partial \bm{\gamma}}\Bigg|_{\bm{\gamma}=\bm{\gamma}^{(t)}} \right)^\top (\bm{\gamma} - \bm{\gamma}^{(t)}). \label{eq:approx_linearization}
\end{equation}
Let $\bar{\mu}_{i}^{(t)}$ denote the first term (the simulated marginal mean), and let $\bar{\bm{D}}_{i}^{(t)}$ denote the bracketed term (the simulated gradient vector, computed as the average of the individual draw derivatives). The expansion simplifies to $\bar{\mu}_i(\bm{\gamma}) \approx \bar{\mu}_i^{(t)} + \bar{\bm{D}}_i^{(t)\top} (\bm{\gamma} - \bm{\gamma}^{(t)})$. This produces the approximate quasi-score equation:
\begin{equation}
    \bar{\bm{U}}(\bm{\gamma}) = \sum_{i=1}^n \bar{\bm{D}}_{i}^{\top}  \bar{W}_{i} (y_i - \bar{\mu}_{i}) = \bm{0}, \label{eq:approx_score}
\end{equation}
where $\bar{W}_{i}=\bar{V}_{i}^{-1}$, and $\bar{V}_{i}$ is the Monte Carlo approximation defined in Eq. \ref{eq:approx_var} replacing the exact variance $V_{i}$. To find the root of Eq. \ref{eq:approx_score}, we execute the TSQLEM algorithm as follows:

\begin{enumerate}
    \item \textbf{Initialization:} We require a stable starting value, $\bm{\gamma}^{(0)}$. We recommend setting $\bm{\gamma}^{(0)}$ to the naive estimates obtained by directly regressing $\bm{y}$ on $\hat{\bm{\eta}}$ using a standard GLM specified with the same target family and link function.
    
    \item \textbf{The E-Step:} At iteration $t$, for each observation $i = 1, \dots, n$, compute the predicted value $\mu_{is}^{(t)}$ for every draw $s = 1, \dots, B$. Calculate $\bar{\mu}_{i}^{(t)}$ by averaging $\mu_{is}^{(t)}$, and find $\bar{\bm{D}}_{i}^{(t)}$.
    
    \item \textbf{Weight Calculation and Delta Correction:} Compute the approximate marginal variance $\bar{V}_{i}^{(t)}$. In a standard IRLS scheme, the quasi-likelihood weight is simply the reciprocal of the variance. However, evaluating $\bar{W}_{i}^{(t)} = 1/\bar{V}_{i}^{(t)}$ directly may introduce finite sample bias, since $\bar{V}_{i}^{(t)}$ is a simulation based estimate. To remove this weight inflation, we apply a second-order Delta Method correction to debias the weights:
    \begin{equation}
       \bar{W}_{i}^{(t)} = \frac{1}{\bar{V}_{i}^{(t)}} - \frac{S^2_{\bar{V}, i}}{\left(\bar{V}_{i}^{(t)}\right)^3}, \label{eq:delta_weights}
    \end{equation}
    where  $S^2_{\bar{V} i}$ is the estimated variance of the variance estimators across the $B$ draws. These corrected weights are subsequently bounded away from zero to ensure numerical stability.
    
    \item \textbf{The M-Step:} We update the structural parameters directly via a robust Fisher Scoring matrix step. Let $\bar{\bm{D}}$ be the $n \times (p+1)$ gradient matrix, and $\bar{\bm{W}}^{(t)}$ be the diagonal weight matrix. Construct the approximate gradient vector $\bar{\bm{U}}(\bm{\gamma}^{(t)})$ and the expected Information matrix $\bar{\bm{H}}(\bm{\gamma}^{(t)}) = \bar{\bm{D}}^\top \bar{\bm{W}}^{(t)} \bar{\bm{D}}$. The parameters are updated as:
    \begin{equation}
        \bm{\gamma}^{(t+1)} = \bm{\gamma}^{(t)} + \bar{\bm{H}}(\bm{\gamma}^{(t)})^{-1} \bar{\bm{U}}(\bm{\gamma}^{(t)}). \label{eq:m_step_update}
    \end{equation}
    
    \item \textbf{Convergence Check:} Evaluate the relative change in the linearized working predictors $\bar{\bm{\eta}}^{(t)} = \bar{\bm{D}} \bm{\gamma}^{(t)}$. Repeat Steps 2--4 until the relative change falls below a predefined tolerance threshold.
\end{enumerate}

The iterative scheme above establishes the MC simulation based framework. For models that possess exact analytical solutions, Eq. \ref{eq:exact_score} is optimized  directly  (no MC error). We demonstrate the mechanics of both pathways below utilizing binary and Gamma distributions, respectively.

\subsection{Binomial Data}

In this section, we demonstrate the mechanics of the TSQLEM algorithm when $y_i \mid \bm{\eta}_i$ follows a Binomial distribution under logit and probit links.

If $y_i \mid \bm{\eta}_i$ follows a Binomial distribution with $m$ independent trials and success probability per trial $g^{-1}(\beta_0 + \bm{\eta}_i^\top \bm{\beta})/m$, then following Equation \ref{eq:cond_var}, the conditional variance is:

\begin{equation}
V_i= \mu_i(1 - \mu_i/m) + \frac{m-1}{m}var\left(g^{-1}(\beta_0 + \bm{\eta}_i^\top \bm{\beta}) \mid \bm{Z}_i\right)
\end{equation}

Consequently, the marginal distribution of $y_i \mid \bm{Z}_i$ exhibits overdispersion and therefore does not follow a standard Binomial distribution when $m > 1$. However, when $m=1$ (the binary case), the variance collapses cleanly to:
\begin{equation}
V_i = var(y_i \mid \bm{Z}_i) = \mu_i (1-\mu_i). \label{eq:varBern}
\end{equation}

The scheme below is described for $m=1$ for simplicity, but it proceeds in the same manner when $m>1$. Let $\bm{\gamma}^{(t)}$ be the parameter estimates at iteration $t$. The iterative scheme is described as follows:

\begin{enumerate}
    \item \textbf{The E-Step:} For each subject $i$ and draw $s = 1, \dots, B$, compute the predicted probability under the logistic link:
    \begin{equation}
        \mu_{is}^{(t)} = \frac{1}{1 + \exp(-\bm{X}_{is}^\top \bm{\gamma}^{(t)})}, \label{eq:logit_prob}
    \end{equation}
    or
   \begin{equation}
        \mu_{is}^{(t)} = \Phi\left(\bm{X}_{is}^\top \bm{\gamma}^{(t)}\right), \label{eq:probit_prob}
    \end{equation}
    under the probit link, where $\Phi(\cdot)$ is the cumulative distribution function for a standard normal distribution.
    
    The simulated marginal mean (the expected probability) is obtained by averaging across the Monte Carlo draws:
    \begin{equation}
        \bar{\mu}_i^{(t)} = \frac{1}{B} \sum_{s=1}^B \mu_{is}^{(t)}. \label{eq:logit_agg_prob}
    \end{equation}
    The derivatives of the logistic and probit inverse link functions with respect to the structural parameters $\bm{\gamma}$ are $\bm{X}_{is} \mu_{is}^{(t)}(1 - \mu_{is}^{(t)})$ and $\bm{X}_{is} \phi\left(\bm{X}_{is}^\top \bm{\gamma}^{(t)}\right)$, respectively, where $\phi(\cdot)$ is the probability density function for a standard normal distribution. Therefore, the simulated gradient vector for subject $i$ evaluates to:
    \begin{equation}
        \bar{\bm{D}}_{i}^{(t)} = \frac{1}{B} \sum_{s=1}^B \bm{X}_{is} \mu_{is}^{(t)} \left(1 - \mu_{is}^{(t)}\right), \label{eq:logit_design}
    \end{equation}
    under the logit link, and 
  \begin{equation}
        \bar{\bm{D}}_{i}^{(t)} = \frac{1}{B} \sum_{s=1}^B \bm{X}_{is} \phi\left(\bm{X}_{is}^\top \bm{\gamma}^{(t)}\right), \label{eq:probit_design}
    \end{equation}
    under the probit link.

    \item \textbf{Weight Calculation:} Applying Equation \ref{eq:approx_var} to the Bernoulli draws yields:
    \begin{equation}
        \bar{V}_i^{(t)} = \frac{1}{B} \sum_{s=1}^B \mu_{is}^{(t)} \left(1 - \mu_{is}^{(t)}\right) + \left[ \frac{1}{B} \sum_{s=1}^B \left(\mu_{is}^{(t)}\right)^2 - \left(\bar{\mu}_{i}^{(t)}\right)^2 \right]. \label{eq:logit_total_var}
    \end{equation}
    The $(\mu_{is}^{(t)})^2$ terms cancel, allowing the total variance to collapse into the variance of the marginal Bernoulli distribution:
    \begin{equation}
        \bar{V}_i^{(t)} = \bar{\mu}_i^{(t)} \left(1 - \bar{\mu}_i^{(t)}\right), \label{eq:logit_marginal_var}
    \end{equation}
    which conforms to the plug-in estimate of the formula given in Eq. \ref{eq:varBern}. Following this, unbiased weights are computed according to Eq. \ref{eq:delta_weights}.
    
    Because this variance is bounded ($0 \le \bar{V}_i^{(t)} \le 0.25$), $S^2_{\bar{V}, i}$ tends to be small. Therefore, the Delta Method correction from Eq. \ref{eq:delta_weights} does not play a crucial role in this case. However, the preceding simplifications do not generally hold for the Binomial family when $m>1$. In such cases, Equation \ref{eq:approx_var} must be utilized, and the Delta Method correction becomes extremely important to properly debias the weights.

    \item \textbf{The M-Step and Convergence:} The structural parameters are updated utilizing the Fisher Scoring matrix step established in Eq. \ref{eq:m_step_update}:
    \begin{equation}
        \bm{\gamma}^{(t+1)} = \bm{\gamma}^{(t)} + \left( \bar{\bm{D}}^\top \bar{\bm{W}}^{(t)} \bar{\bm{D}} \right)^{-1} \bar{\bm{D}}^\top \bar{\bm{W}}^{(t)} \left( \bm{y} - \bar{\bm{\mu}}^{(t)} \right). \label{eq:logit_update}
    \end{equation}
    The algorithm terminates when the relative change in the linearized working predictors, $\bar{\bm{\eta}}^{(t)} = \bar{\bm{D}} \bm{\gamma}^{(t)}$, falls below a predefined tolerance threshold:
    \begin{equation}
        \sum_{i=1}^n \left( \bar{\eta}_i^{(t+1)} - \bar{\eta}_i^{(t)} \right)^2 < \text{tol} \times \sum_{i=1}^n \left( \bar{\eta}_i^{(t)} \right)^2,  \label{eq:convergence}
    \end{equation}
    where $\text{tol}$ is typically set to $10^{-4}$.
\end{enumerate}

\subsubsection{Bernoulli Probit Data}

The MC integration scheme we have just described generally applies even when the exact moments can be found. As demonstrated previously, both the conditional mean and variance can be found analytically under the probit link when $m=1$. Utilizing those exact moments as detailed next is highly attractive due to computational efficiency, unless one suspects the deviation from normality is large and the draws generated in Stage 1 were not taken from a multivariate normal distribution. 

The optimization relies  on the exact marginal quasi-score equation (Eq. \ref{eq:exact_score}) and proceeds as follows:
\begin{enumerate}

\item \textbf{Exact Moments and Gradient:} For each subject $i$, compute the exact marginal mean $\mu_i^{(t)}$ utilizing Eq. \ref{eq:mean_probit} evaluated at $\bm{\gamma}^{(t)}$. Let $s_i = \sqrt{1 + \sigma_{\eta,i}^2}$ be the scaling factor. Applying the chain and quotient rules, the gradient vector $\bm{D}_i^{(t)} = \frac{\partial\mu_i(\bm{\gamma})}{\partial\bm{\gamma}}\big|_{\bm{\gamma}=\bm{\gamma}^{(t)}}$ evaluates to:
\begin{equation}
\bm{D}_i^{(t)} = \phi\left(\frac{\bm{X}_{obs,i}^\top \bm{\gamma}^{(t)}}{s_i}\right) \left( \frac{\bm{X}_{obs,i}}{s_i} - \frac{(\bm{X}_{obs,i}^\top \bm{\gamma}^{(t)})(\tilde{\bm{\Sigma}}_i \bm{\gamma}^{(t)})}{s_i^3} \right),
\end{equation}
where $\phi(\cdot)$ is the standard normal probability density function.

\item \textbf{Weight Calculation:} The exact marginal variance $V_i^{(t)}$ is fully realized via Eq. \ref{eq:varBern}. The weights are given by:
\begin{equation}
W_i^{(t)} = \left(V_i^{(t)}\right)^{-1} = \frac{1}{\mu_i^{(t)}(1 - \mu_i^{(t)})}.
\end{equation}

\item \textbf{The M-Step and Convergence:} The structural parameters are updated by seeking the root of the exact marginal quasi-score equation (Eq. \ref{eq:exact_score}). Unlike the approximate score equation (Eq. \ref{eq:approx_score}), which relies on simulated gradients, we utilize the exact derived quantities:
\begin{equation}
U(\bm{\gamma}) = \sum_{i=1}^{n} {\bm{D}_i^{(t)}}^\top W_i^{(t)} (y_i - \mu_i^{(t)}) = 0.
\end{equation}
Let $\bm{D}$ be the $n \times (p+1)$ gradient matrix, and $\bm{W}^{(t)}$ be the diagonal weight matrix. We construct the expected Information matrix $\bm{H}(\bm{\gamma}^{(t)}) = \bm{D}^\top \bm{W}^{(t)} \bm{D}$. The parameters are updated via the Fisher Scoring step:
\begin{equation}
\bm{\gamma}^{(t+1)} = \bm{\gamma}^{(t)} + \left(\bm{D}^\top \bm{W}^{(t)} \bm{D}\right)^{-1} \bm{D}^\top \bm{W}^{(t)} (\bm{y} - \bm{\mu}^{(t)}).
\end{equation}
The algorithm terminates as above.
\end{enumerate}

Finally, for binomial probit models, the exact conditional mean and gradient can be found analytically regardless of the value of $m$, unlike the variance which becomes intractable for $m > 1$. Instead of resorting to fully Monte Carlo integration, and in order to avoid introducing unnecessary stochastic noise into the parameter updates, we recommend a hybrid procedure. We proceed using the exact analytical mean and gradient as just described, but estimate the de-biased weights via Monte Carlo integration as detailed in the previous section. The structural parameters are then updated using:
\begin{equation}
\bm{\gamma}^{(t+1)} = \bm{\gamma}^{(t)} + \left(\bm{D}^\top \bar{\bm{W}}^{(t)} \bm{D}\right)^{-1} \bm{D}^\top \bar{\bm{W}}^{(t)} (\bm{y} - \bm{\mu}^{(t)}).
\end{equation}

\subsection{Gamma Data}

The TSQLEM algorithm extends beyond binary outcomes to other distributions in the exponential family. Research frequently requires modeling count data (e.g., Poisson and Negative Binomial) or  positive right skewed data (Gamma). While some traditional programs support count data, they rely heavily on numerical integration algorithms that scale poorly as the number of latent variables increases. Furthermore, estimation for Gamma distributions within latent variable frameworks is rarely supported. The TSQLEM method  resolves these limitations, accommodating virtually any generalized linear model family.

In this section, we demonstrate the TSQLEM algorithm when $y_i \mid \bm{\eta}_i$ follows a Gamma distribution with a log link function. A very similar development applies under the Poisson and Negative Binomial families. As explained previously, the first and second moments of $y_i \mid \bm{Z}_i$ can be found analytically under the log link. 

For the Gamma family, the variance is proportional to the square of the  mean: $V(\beta_0 + \bm{\eta}_i^\top \bm{\beta}, \phi) = \phi \left[\exp(\beta_0 + \bm{\eta}_i^\top \bm{\beta})\right]^2$, where $\phi$ is the dispersion parameter. Following Eq. \ref{eq:cond_var}, the exact marginal variance decomposes as:
\begin{align}
V_i = \mathrm{var}(y_i \mid \bm{Z}_i) &= E\left[ \phi \exp\left(2(\beta_0 + \bm{\eta}_i^\top \bm{\beta})\right) \mid \bm{Z}_i \right] + \mathrm{var}\left( \exp(\beta_0 + \bm{\eta}_i^\top \bm{\beta}) \mid \bm{Z}_i \right). \label{eq:gamma_var_decomp}
\end{align}

Utilizing the properties of the log-normal distribution, the expectation of the squared inverse link evaluates precisely to $E[\exp(2(\beta_0 + \bm{\eta}_i^\top \bm{\beta})) \mid \bm{Z}_i] = \mu_i^2 \exp(\sigma_{\eta, i}^2)$, where $\mu_i$ is the exact marginal mean (Eq. \ref{eq:exact_mean}) and $\sigma_{\eta, i}^2 = \bm{\gamma}^\top \tilde{\bm{\Sigma}}_i \bm{\gamma}$. Therefore, the total variance simplifies to the exact analytical form established in Eq. \ref{eq:exact_var}:
\begin{equation}
V_i = \phi \mu_i^2 \exp(\sigma_{\eta, i}^2) + \mu_i^2 \left( \exp(\sigma_{\eta, i}^2) - 1 \right). \label{eq:gamma_exact_var}
\end{equation}

The algorithm proceeds as follows:

\begin{enumerate}
    \item \textbf{Exact Moments and Gradient:} For each subject $i$, compute the exact analytical marginal mean $\mu_i^{(t)}$ utilizing Eq. \ref{eq:exact_mean} evaluated at $\bm{\gamma}^{(t)}$. 
    With $\mu_i^{(t)} = \exp\left( \bm{X}_{obs, i}^\top \bm{\gamma}^{(t)} + \frac{1}{2} \bm{\gamma}^{(t)\top}\tilde{\bm{\Sigma}}_i \bm{\gamma}^{(t)} \right)$, the gradient vector evaluates to:
    \begin{equation}
        \bm{D}_{i}^{(t)} = \mu_i^{(t)} \left( \bm{X}_{obs, i} + \tilde{\bm{\Sigma}}_i \bm{\gamma}^{(t)} \right). \label{eq:gamma_design}
    \end{equation}

    \item \textbf{Dispersion Estimation and Weight Calculation:} Unlike the Bernoulli distribution, the Gamma marginal variance (Eq. \ref{eq:gamma_exact_var}) depends on an unknown dispersion parameter $\phi$. This parameter is estimated at each iteration using a profile likelihood approach. Specifically, we formulate a Gaussian working model and minimize its negative log-likelihood with respect to $\log(\phi)$, incorporating both the distributional dispersion and the latent variance derived analytically to ensure the dispersion estimate accounts for measurement uncertainty:
    \begin{equation}
        l(\phi) = \sum_{i=1}^n \log(V_i(\phi)) + \sum_{i=1}^n \frac{(y_i - \mu_i^{(t)})^2}{V_i(\phi)}, \label{eq:gamma_profile}
    \end{equation}
    where $V_i(\phi)$ relies on the current analytical moments evaluated at $\bm{\gamma}^{(t)}$. Upon obtaining $\hat{\phi}^{(t)}$,  $V_i^{(t)}$ is fully realized via Eq. \ref{eq:gamma_exact_var}, and $
        W_{i}^{(t)} = \left( V_i^{(t)} \right)^{-1}$.

    \item \textbf{The M-Step and Convergence:} The structural parameters are updated by seeking the root of the exact marginal quasi-score equation (Eq. \ref{eq:exact_score}) with termination criterion identical to the binary case.
    
\end{enumerate}

\subsection{The Full Conditional Distribution of the Latent Space}

Once the TSQLEM algorithm converges and the structural parameters $(\hat{\bm{\gamma}}, \hat{\phi})$ are obtained, one can estimate the full conditional distribution $f(\bm{\eta}_i \mid \bm{Z}_i, y_i)$.

Given the conditional independence of $y_i$ and $\bm{Z}_i$ given $\bm{\eta}_i$, the conditional distribution of the latent variables given all observed data evaluates to:
\begin{align}
    f(\bm{\eta}_i \mid \bm{Z}_i, y_i) &= \frac{f(y_i \mid \bm{\eta}_i, \bm{Z}_i) f(\bm{\eta}_i \mid \bm{Z}_i)}{f(y_i \mid \bm{Z}_i)} \nonumber \\
    &\propto f(y_i \mid \bm{\eta}_i ; \hat{\bm{\gamma}}, \hat{\phi}) f(\bm{\eta}_i \mid \bm{Z}_i ; \hat{\bm{\eta}}_i, \bm{\Sigma}_i), \label{eq:full_conditional}
\end{align}
which is proportional to the product of the Stage 2 structural density and the Stage 1 measurement density. 

Typically, this target distribution does not correspond to a known functional form (e.g., combining a strictly binary probit structural outcome with a multivariate normal measurement approximation). To draw samples from this unnormalized target distribution, we implement a Metropolis-Hastings (MH) algorithm as follows:

\begin{enumerate}
    \item \textbf{Initialization:} For each subject $i$, initialize the Markov chain at the Stage 1 empirical Bayes estimate: $\bm{\eta}_i^{(0)} = \hat{\bm{\eta}}_i$.
    
    \item \textbf{Proposal Step:} At iteration $s$, draw a candidate vector $\bm{\eta}_i^*$ from  $\bm{\eta}_i^* \sim \mathcal{N}(\bm{\eta}_i^{(s-1)}, c\bm{\Sigma}_i)$, where $c$ is a tuning scalar optimized to achieve a desirable acceptance rate (e.g., $0.23$ to $0.44$).
    
    \item \textbf{Acceptance Probability:} Calculate the MH ratio comparing the target density evaluated at the candidate state versus the current state. Utilizing Eq. \ref{eq:full_conditional}, the acceptance probability evaluates to:
    \begin{equation}
        \alpha = \min \left( 1, \frac{f(y_i \mid \bm{\eta}_i^* ; \hat{\bm{\gamma}}, \hat{\phi}) f(\bm{\eta}_i^* \mid \bm{Z}_i ; \hat{\bm{\eta}}_i, \bm{\Sigma}_i)}{f(y_i \mid \bm{\eta}_i^{(s-1)} ; \hat{\bm{\gamma}}, \hat{\phi}) f(\bm{\eta}_i^{(s-1)} \mid \bm{Z}_i ; \hat{\bm{\eta}}_i, \bm{\Sigma}_i)} \right). \label{eq:mh_ratio}
    \end{equation}
    
    \item \textbf{Update:} Draw a uniform random variable $u \sim \mathrm{U}(0, 1)$. If $u < \alpha$, accept the candidate and set $\bm{\eta}_i^{(s)} = \bm{\eta}_i^*$. Otherwise, reject the candidate and retain the current state: $\bm{\eta}_i^{(s)} = \bm{\eta}_i^{(s-1)}$.
\end{enumerate}

Following an adequate burn-in period to allow the chain to reach stationarity, the retained draws form the empirical distribution of $f(\bm{\eta}_i \mid \bm{Z}_i, y_i)$. 

\section{Simulation Study}

To evaluate the TSQLEM method, a comprehensive simulation study was conducted comparing its performance against several alternative estimation frameworks described later in this section.

\subsection{Simulation Design and Data Generation}

The simulation maintains fixed parameters across all conditions: sample size $n = 4000$, $p = 10$ latent variables, and $J = 40$ indicators (4 items per latent variable). The true latent correlation matrix $\bm \Phi$ features off-diagonal values drawn uniformly between $0.40$ and $0.70$. The factor loadings $\Lambda$ range from $0.50$ to $0.85$. The latent variables are associated with a binary outcome $y$ via a probit link with an intercept $\beta_0 = -0.3$. Data generation varies across three factors (a $2 \times 2 \times 2$ factorial design) resulting in 8 different simulation scenarios:

(1) \textit{Latent Variable Distribution}, generated either as Gaussian ($\bm \eta \sim N(\bm 0, \bm\Phi)$) or as $\chi^2_3$ distribution (centered and scaled) produced via the Normal to Anything approach to approximately preserve the target correlation matrix;
(2) \textit{Indicator Distributions}, comparing a symmetric baseline (1 normal continuous, 1 symmetric binary, and 2 symmetric 5-category ordinal items per latent variable) against a non-symmetric setting where latent variables 6--10 utilize a Gamma distributed item (centered and scaled), an asymmetric 4-category ordinal item, and two asymmetric 5-category ordinal items; and
(3) \textit{Structural Slope Specification}, contrasting a distributed profile ($\beta = \{0.7, -0.6, 0.9, 0.5, -0.4, 0.6, -0.7, 0.5, -0.4, 0.8\}$) against a concentrated profile ($\beta = \{0.1, -0.15, 0.05, 0.1, -0.1, 1.2, -1.3, 1.0, -1.1, 1.25\}$).

  
The baseline scenario is formed under the Gaussian latent space, symmetric indicators, and distributed structural slopes. Each scenario is evaluated across 500 independent replications. All scenarios are listed in Table 1.

\begin{table}[h]
\centering
\caption{Simulation Design Scenarios (S)}
\label{tab:simulation_scenarios}
\begin{tabular}{lllc}
\hline
\textbf{S} & \textbf{Latent} & \textbf{Indicator} & \textbf{Slope} \\ \hline
1 & Gaussian & Symmetric & Distributed \\
2 & $\chi_3^2$ & Symmetric & Distributed \\
3 & Gaussian & Asymmetric & Distributed \\
4 & $\chi_3^2$ & Asymmetric & Distributed \\
5 & Gaussian & Symmetric & Concentrated \\
6 & $\chi_3^2$ & Symmetric & Concentrated \\
7 & Gaussian & Asymmetric & Concentrated \\
8 & $\chi_3^2$ & Asymmetric & Concentrated \\ \hline
\end{tabular}
\end{table}

\subsection{Methods and Results}

For the TSQLEM approach, continuous ML CFA approximation is utilized (treating discrete items as continuous) to approximate the conditional distribution of $\bm{\eta} \mid \bm{Z}$ as explained in Section 2. Having said that, for categorical indicators, measurement precision is heteroscedastic. A theoretically superior extension of TSQLEM would involve extracting ($\Sigma_i$) for each subject. However, extracting $\Sigma_i$ for high-dimensional models is computationally demanding.

The performance of the TSQLEM method is compared against the following estimation frameworks implemented via their respective R packages:

\begin{enumerate}
    \item \textbf{FIML (\texttt{OpenMx}):}  Evaluated via the \texttt{OpenMx} package (Boker et al., 2011). Unlike MML, FIML integrates over the observed variables. For continuous data, this relies on highly efficient multivariate normal density functions. However, when the indicators are categorical, FIML must compute the probability of a response pattern by evaluating a multidimensional multivariate normal cumulative distribution function bounded by specific thresholds. In this case, applying FIML to a 40-item model requires intractable multidimensional integrations. To facilitate FIML computations, all measurement indicators were considered as continuous, as in stage 1 of TSQLEM.
    
    \item \textbf{WLSMV (\texttt{lavaan}):}  WLSMV operates by calculating the asymptotic polychoric, tetrachoric, and polyserial correlation matrices of the indicators, and subsequently applying diagonally weighted least squares for parameter optimization. Utilizing \texttt{lavaan} R package (Rosseel, 2012), four WLSMV estimators are considered depending on how indicators are treated: (1) all discrete indicators treated as ordered (WLSMV245); (2) binary and 4-category indicators treated as ordered (WLSMV24); (3) only binary indicators treated as ordered (WLSMV2); and (4) all indicators treated as continuous (WLSMV). 
\end{enumerate}

The results are shown in Table 2, presented in terms of the root mean squared error (RMSE) and the root mean squared bias (RMSB) for the structural slopes. The TSQLEM and FIML methods achieved a 100\% convergence rate across all scenarios. In contrast, the convergence of the WLSMV family varied drastically depending on the treatment of the categorical indicators. For example, when all discrete indicators were treated as ordered (WLSMV245), convergence was perfect (100\%) in Scenario 2, but decreased to 8\% in Scenario 4. When the WLSMV estimator treated all indicators as continuous, convergence rates stabilized somewhat but still fluctuated, achieving 89\% in Scenario 1 and 99\% in Scenario 8.

TSQLEM demonstrated stable and consistent performance across all simulation scenarios. TSQLEM yielded the  lowest RMSE in 7 scenarios, and the lowest RMSB in 5 scenarios, including the baseline case where we expect the simultaneous estimators to dominate. TSQLEM achieved this stability despite the deliberate degree of misspecification at the measurement level introduced by treating categorical indicators as continuous. When the latent distribution deviated from normality (Scenarios 2, 4, 6, and 8), TSQLEM maintained lower structural RMSE.

The performance of the WLSMV estimators was highly unstable. For instance, the WLSMV24 estimator (treating binary and 4-category indicators as ordered) performed exceptionally well in Scenario 7, yielding the lowest RMSE and RMSB of all evaluated methods. However, that same estimator utterly failed in Scenarios 4 and 8.

Across all simulation scenarios and based on separate $15$ replications, the median computation time  per replication for the TSQLEM framework was $0.10$ seconds ( max: $0.6$ seconds). In contrast, the median times for the FIML and various WLSMV estimators were $17.3$ seconds (max: $27.7$ seconds) and $28.2$ seconds (max: $3136.7$ seconds), respectively.

\begin{table}[h]
\centering
\caption{Simulation Results: Root Mean Squared Error (RMSE) and Root Mean Squared Bias (RMSB) for Structural Slope Estimates.}
\label{tab:sim_slopes}
\resizebox{\textwidth}{!}{
\begin{tabular}{lcccccc}
\hline
\textbf{S} & \textbf{TSQLEM} & \textbf{WLSMV245} & \textbf{WLSMV24} & \textbf{WLSMV2} & \textbf{WLSMV} & \textbf{FIML} \\ \hline
1 & \textbf{0.181} (\textcolor{red}{\textbf{0.016}}) & 0.190 (\textcolor{red}{0.025}) & 0.190 (\textcolor{red}{0.025}) & 0.190 (\textcolor{red}{0.025}) & 0.185 (\textcolor{red}{0.019}) & 0.354 (\textcolor{red}{0.074}) \\
2 & \textbf{0.168} (\textcolor{red}{0.031}) & 0.182 (\textcolor{red}{\textbf{0.017}}) & 0.181 (\textcolor{red}{0.031}) & 0.181 (\textcolor{red}{0.031}) & 0.179 (\textcolor{red}{0.036}) & 0.173 (\textcolor{red}{0.021}) \\
3 & \textbf{0.137} (\textcolor{red}{0.070}) & 13.667 (\textcolor{red}{0.539}) & 0.228 (\textcolor{red}{0.089}) & 0.149 (\textcolor{red}{0.056}) & 0.144 (\textcolor{red}{\textbf{0.051}}) & 0.189 (\textcolor{red}{0.149}) \\
4 & \textbf{0.131} (\textcolor{red}{\textbf{0.071}}) & 13.978 (\textcolor{red}{3.561}) & 59.385 (\textcolor{red}{2.676}) & 0.209 (\textcolor{red}{0.165}) & 20.946 (\textcolor{red}{0.946}) & 0.279 (\textcolor{red}{0.260}) \\
5 & \textbf{0.212} (\textcolor{red}{\textbf{0.023}}) & 0.226 (\textcolor{red}{0.041}) & 0.230 (\textcolor{red}{0.041}) & 0.230 (\textcolor{red}{0.041}) & 0.221 (\textcolor{red}{0.027}) & 0.224 (\textcolor{red}{0.059}) \\
6 & \textbf{0.199} (\textcolor{red}{\textbf{0.107}}) & 0.205 (\textcolor{red}{0.110}) & 0.208 (\textcolor{red}{0.117}) & 0.208 (\textcolor{red}{0.117}) & 0.212 (\textcolor{red}{0.126}) & 0.213 (\textcolor{red}{0.131}) \\
7 & 0.256 (\textcolor{red}{0.226}) & 16.354 (\textcolor{red}{1.054}) & \textbf{0.246} (\textcolor{red}{\textbf{0.054}}) & 0.315 (\textcolor{red}{0.293}) & 0.313 (\textcolor{red}{0.293}) & 0.371 (\textcolor{red}{0.358}) \\
8 & \textbf{0.237} (\textcolor{red}{\textbf{0.215}}) & 65.787 (\textcolor{red}{20.769}) & 23.122 (\textcolor{red}{1.064}) & 16.390 (\textcolor{red}{0.824}) & 0.390 (\textcolor{red}{0.380}) & 0.473 (\textcolor{red}{0.468}) \\ \hline
\end{tabular}
}
\end{table}

\subsection{Logistic Regression}

While WLSMV and FIML evaluated above are restricted to the probit link, the TSQLEM approach can accommodate any differentiable link function, including the logit link. To evaluate the framework's Monte Carlo integration scheme (detailed in Section 2.2.1; with the number of random draws $B=3000$), we compared TSQLEM against Marginal Maximum Likelihood evaluated via the Metropolis-Hastings Robbins-Monro algorithm (MML-MHRM), implemented in the R package mirt (Chalmers, 2012), which utilizes the logit link. 

The MHRM algorithm becomes computationally exhaustive when $p=10$. Therefore, this comparison was restricted to the baseline scenario with $300$ replications to ensure computational feasibility for the MHRM estimator. The outcome $y$ was generated via the logit link. To produce results that are similar in magnitude to Table 2, the true regression coefficients in Table 1  are divided by $1.7$ and the resulting estimates from both approaches are multiplied by $1.7$. To accommodate the graded response model in the R-package mirt, the continuous indicators were converted to ordinal variables with 5 categories before initiating the estimation procedures. 

Despite the misspecification induced in stage 1 of the TSQLEM approach, it has largely outperformed the MML-MHRM in terms of bias (RMSB of $0.009$ vs $0.043$) in this ideal baseline scenario where MML-MHRM's parametric assumptions are perfectly met. The RMSEs for both approaches are indistinguishable ($0.177$ vs $0.172$). 

This reduction in bias was achieved alongside significant computational gains. Based on 10 replications,  the median computation time per replication for our approach is $43.8$ seconds (max: $114.5$ seconds) compared to $1167.7$ seconds (max : $1715.3 $ seconds) for MML-MHRM.

\subsection{Poisson GSEM}

As described in Section 2, the TSQLEM approach accommodates all standard and quasi-likelihood families within the generalized linear modeling framework. To demonstrate its usage for count data, we applied the method to the eight simulation scenarios outlined in Table 1 for a structural outcome that follows a Poisson distribution with a $\log$ link. The regression coefficients were multiplied by $0.75$ to prevent extreme exponentiation of the simulated counts. All other simulation parameters remained identical. The results are displayed in Table 3.

 Utilizing the $\log$ link allows for analytical solutions. Should the conditional distribution of the latent space depart substantially from normality, Monte Carlo integration can be utilized, provided simulation draws approximating the conditional space can be generated. Here we assumed normality, even under scenarios where it was explicitly violated, and proceeded with the analytical solution to test TSQLEM robustness to this departure. Computationally, a single run requires only a fraction of a second.

We contrasted our method against a naive approach that treats the expected value of the conditional latent space as fixed covariates. As expected, this approach suffered from attenuation bias, and our method significantly reduced this bias. Other approaches that can handle Poisson data include MML in Mplus and Stata, which realistically cannot be performed beyond $p=3$. Likewise, the full Bayesian MCMC estimators remain computationally prohibitive for high dimensional models.

\begin{table}[h]
\centering
\caption{Simulation Results for Poisson GSEM: RMSE (RMSB).}
\label{tab:sim_slopes_poisson}
\begin{tabular}{lcc}
\hline
\textbf{S} & \textbf{TSQLEM} & \textbf{Naive EB} \\ \hline
1 & \textbf{0.112} (\textcolor{red}{\textbf{0.009}}) & 0.188 (\textcolor{red}{0.024}) \\
2 & \textbf{0.619} (\textcolor{red}{\textbf{0.131}}) & 1.250 (\textcolor{red}{0.552}) \\
3 & \textbf{0.122} (\textcolor{red}{\textbf{0.105}}) & 0.179 (\textcolor{red}{0.152}) \\
4 & \textbf{0.148} (\textcolor{red}{\textbf{0.130}}) & 0.657 (\textcolor{red}{0.432}) \\
5 & \textbf{0.124} (\textcolor{red}{\textbf{0.012}}) & 0.190 (\textcolor{red}{0.022}) \\
6 & \textbf{0.865} (\textcolor{red}{\textbf{0.080}}) & 1.216 (\textcolor{red}{0.445}) \\
7 & \textbf{0.189} (\textcolor{red}{\textbf{0.170}}) & 0.255 (\textcolor{red}{0.226}) \\
8 & \textbf{0.135} (\textcolor{red}{\textbf{0.112}}) & 0.557 (\textcolor{red}{0.170}) \\ \hline
\end{tabular}
\end{table}

\section{Application}

The TSQLEM framework was applied to the Big Five Inventory (BFI) dataset to estimate a high-dimensional Generalized Structural Equation Model. The measurement model consists of 25 indicators scored on a 6-point ordinal scale, with five indicators assessing each of the five correlated latent personality traits: Agreeableness, Conscientiousness, Extraversion, Neuroticism, and Openness. In the structural model, these five latent traits serve as simultaneous predictors of a binary outcome (gender). After deletion of cases with missing data, the final dataset consisted of 2,436 respondents.

For this empirical application, Table 4 displays the structural parameter estimates and associated standard errors across several methods, grouped by their respective link functions. A nonparametric bootstrap procedure with 1,000 replications was utilized to calculate the standard errors for the TSQLEM framework.

We initially attempted to estimate the 5-dimensional model using the traditional MML approach, first via the standard Expectation-Maximization (EM) algorithm and  then via Quasi-Monte Carlo Expectation-Maximization (QMCEM). As anticipated, these approaches were unstable due to $p=5$, as the likelihood was decreasing near the maximum likelihood estimate.

The first tier of the table evaluates models utilizing a probit link. In addition to the proposed TSQLEM framework, we include the WLSMV estimator, which treats all indicators as ordinal. With 25 items on a 6-point scale, this model estimates 125 thresholds and requires the computation of 300 contingency tables. At a sample size of n=2436, this sparsity frequently triggers numerical instability. Additionally, we include two estimators that treat all indicators as continuous, which is a reasonable choice given the 6-point scale, implemented via WLSMV and FIML.

The second tier of the table provides results for models utilizing a logit link function, specifically comparing the TSQLEM framework with MML evaluated via the MHRM algorithm. Both MML-MHRM and the ordinal WLSMV account for the true ordinal nature of the 25 indicators.

\begin{table}[!ht]
\centering
\small
\renewcommand{\arraystretch}{0.85}
\setlength{\tabcolsep}{4pt}
\caption{Structural Parameter Estimates and Standard Errors (in parenthesis) for the BFI Application.}
\label{tab:empirical_results}
\begin{tabular}{lcccc}
\toprule
\multicolumn{5}{c}{\textbf{Tier 1: Probit Link Models}} \\
\midrule
\textbf{Predictor} & \textbf{TSQLEM} & \shortstack{\textbf{WLSMV} \\ \textbf{(Ordinal)}} & \shortstack{\textbf{WLSMV} \\ \textbf{(Cont.)}} & \shortstack{\textbf{FIML} \\ \textbf{(Cont.)}} \\
\midrule
Intercept         & 0.48  & 0.50  & 0.49  & 0.48 \\
                  & (0.03) & (0.03) & (0.03) & (0.03) \\[0.5ex]
Agreeableness     & $-$0.36 & $-$0.42 & $-$0.36 & $-$0.37 \\
                  & (0.05) & (0.06) & (0.05) & (0.05) \\[0.5ex]
Conscientiousness & 0.16  & 0.16  & 0.15  & 0.16 \\
                  & (0.04) & (0.04) & (0.04) & (0.04) \\[0.5ex]
Extraversion      & $-$0.05 & $-$0.06 & $-$0.08 & $-$0.05 \\
                  & (0.06) & (0.06) & (0.06) & (0.06) \\[0.5ex]
Neuroticism       & 0.28  & 0.33  & 0.30  & 0.28 \\
                  & (0.04) & (0.04) & (0.04) & (0.04) \\[0.5ex]
Openness          & $-$0.29 & $-$0.32 & $-$0.29 & $-$0.29 \\
                  & (0.04) & (0.05) & (0.04) & (0.04) \\ 
\midrule
\multicolumn{5}{c}{\textbf{Tier 2: Logit Link Models}} \\
\midrule
\textbf{Predictor} & \textbf{TSQLEM} & \shortstack{\textbf{MML-MHRM} \\ \textbf{(Ordinal)}} & & \\
\midrule
Intercept         & 0.79  & 0.81  & & \\
                  & (0.05) & (0.05) & & \\[0.5ex]
Agreeableness     & $-$0.59 & $-$0.73 & & \\
                  & (0.10) & (0.12) & & \\[0.5ex]
Conscientiousness & 0.27  & 0.25  & & \\
                  & (0.07) & (0.07) & & \\[0.5ex]
Extraversion      & $-$0.08 & $-$0.02 & & \\
                  & (0.10) & (0.12) & & \\[0.5ex]
Neuroticism       & 0.46  & 0.49  & & \\
                  & (0.07) & (0.08) & & \\[0.5ex]
Openness          & $-$0.48 & $-$0.51 & & \\
                  & (0.08) & (0.06) & & \\ 
\bottomrule
 \\
\end{tabular}
\end{table}

Notice that FIML and TSQLEM produce almost identical estimates and standard errors in this empirical application, despite employing different statistical frameworks. For this application, both methods treated the homogeneous, 6-point ordinal indicators as continuous variables. Importantly, for binary data under a probit link—assuming the approximate distribution of $\bm {\eta}|\bm {Z}$ is close to multivariate normal—our method produces exact marginal likelihood estimates. The alignment between the TSQLEM and FIML standard errors demonstrates that the two-stage modularization achieves this accuracy without sacrificing statistical efficiency.

The continuous WLSMV estimator yielded structural parameters that are highly similar to both TSQLEM and FIML, with the exception of a single minor deviation. This coefficient is not statistically significant, rendering the difference practically negligible. When WLSMV was applied treating the indicators as ordinal, we observed a slight inflation across the parameter estimates.

When evaluating the models specified with a logit link, the largest disparities between the TSQLEM framework and the MML-MHRM estimator occur in the Agreeableness and Extraversion coefficients. The difference in Extraversion is inconsequential, as the coefficient does not reach statistical significance under either estimation method. For Agreeableness, the MHRM estimate deviates not only from our TSQLEM framework but from all other evaluated methods. Generally, dividing a logit link coefficient by a scaling factor of 1.7 yields a close approximation of its probit link equivalent. When applying this conversion, the TSQLEM logit estimate aligns with the probit coefficients produced by TSQLEM, FIML, and continuous WLSMV, whereas applying this scaling to the MHRM estimate does not yield a similar alignment.

To demonstrate the framework’s ability to extract the full conditional distribution of the latent space (Section 2.6), we utilized the Metropolis-Hastings MCMC algorithm to generate individual level latent distributions. Figure 1 illustrates a counterfactual scenario for a single selected respondent. Holding the respondent's 25-item measurement profile  constant, we extracted the latent densities under two structural conditions: y = 0 and y = 1. The divergence between the distributions in Agreeableness and Openness reflects the non-linear updating effect of the structural model. For traits with non-significant structural effects (e.g., Extraversion), the posterior densities perfectly overlap.

The visual divergence between the distributions in Figure 1 appears modest in absolute magnitude, since the Stage 2 structural update does not overwrite the measurement model (derived from a 25-item profile), but rather applies a marginal correction based on a single binary outcome.

\begin{figure}[htbp]
    \centering
    
    \includegraphics[width=0.95\textwidth]{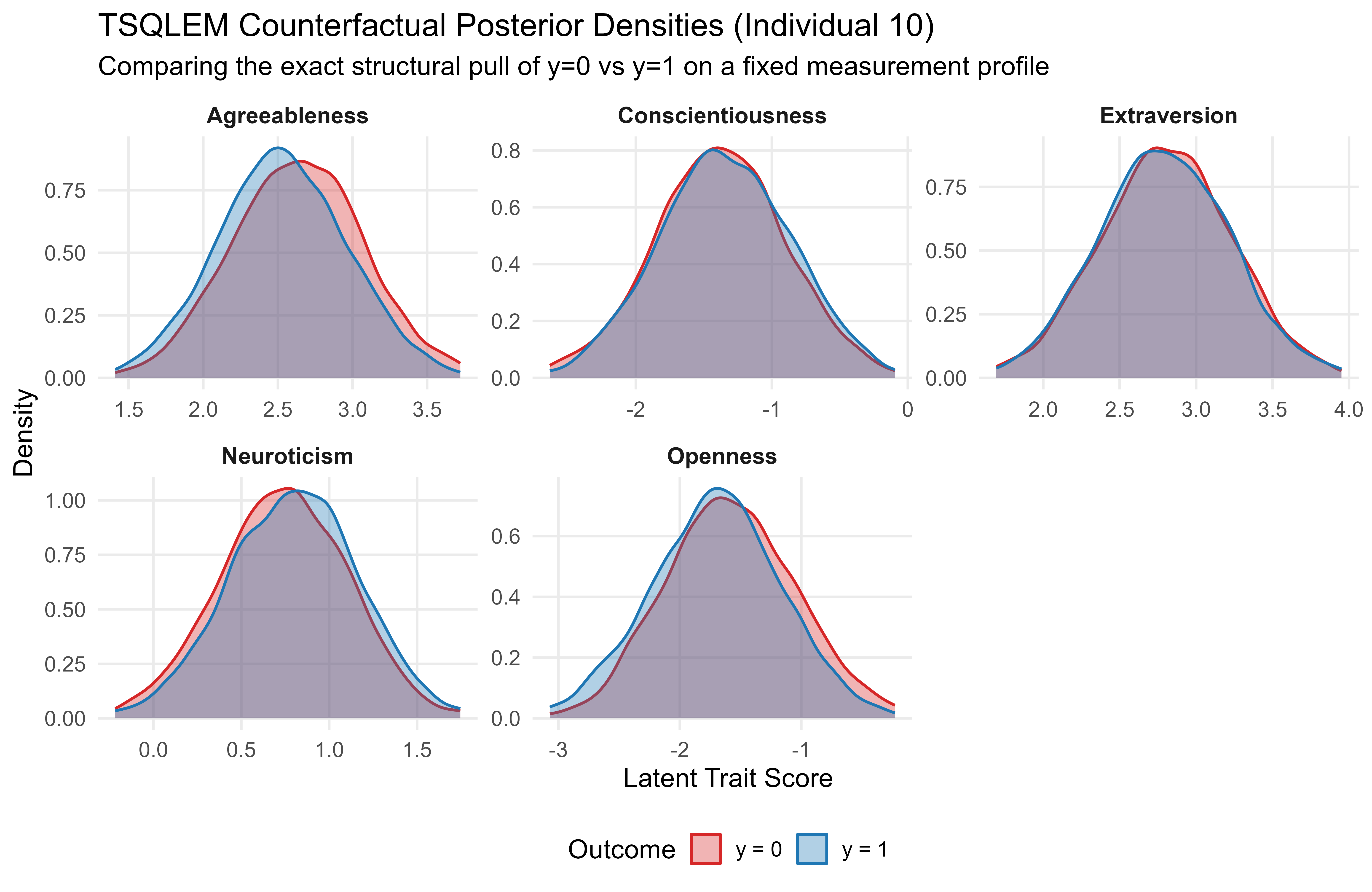}
    \caption{Counterfactual conditional posterior density distributions of the five latent traits for a single respondent, extracted via TSQLEM. The red and blue distributions represent the exact marginal updating effect of the structural model assuming outcomes of $y=0$ and $y=1$, respectively, while holding the empirical measurement profile constant.}
    \label{fig:counterfactual}
\end{figure}

\section{Summary}

 We introduced a TSQLEM framework to address the computational demands and numerical instabilities of estimating high dimensional Generalized Structural Equation Models. It consists of two stages. The first stage focuses on approximating the latent conditional distribution given the observed indicators utilizing the measurement model in isolation from the structural model. Based on this conditional distribution, the second stage only derives the first and the second moments of the distribution of the outcome variable given the observed indicators to avoid computing the intractable full distribution. Depending on the nature of the link function, those quantities can be found exactly or approximated using  Monte Carlo integration. Subsequently, an IRLS scheme is devised to estimate the structural parameters. Furthermore, the updated latent conditional distribution given the outcome variable can be sampled. 
 
   Extensive simulations demonstrated that TSQLEM achieved stable performance, significantly reduced computational time, and reduced the mean squared error and bias in most scenarios as compared to the FIML and limited-information estimators, and MML-MHRM. Its practical utility was demonstrated using the Big Five Inventory dataset. Unlike existing approaches, this method is highly flexible in that it can accommodate various outcome distributions and link functions as well as large sample sizes and high dimensional latent spaces.

\section*{Software}
 An R package implementing the proposed methodology, TSQLEM, is publicly available on CRAN. The Big Five Inventory empirical analysis in Section 4 can be reproduced using the package's examples.


\begin{thebibliography}{}

\bibitem[Anderson \& Gerbing(1988)]{anderson1988}
Anderson, J. C., \& Gerbing, D. W. (1988). Structural equation modeling in practice: A review and recommended two-step approach. \emph{Psychological Bulletin}, 103(3), 411--423.

\bibitem[Arbuckle(1996)]{arbuckle1996}
Arbuckle, J. L. (1996). Full information estimation in the presence of incomplete data. \emph{Advanced structural equation modeling}, 243--277.

\bibitem[Arminger \& Muthén(1998)]{arminger1998}
Arminger, G., \& Muthén, B. O. (1998). A Bayesian approach to nonlinear latent variable models using the Gibbs sampler and the Metropolis-Hastings algorithm. \emph{Psychometrika}, 63(3), 271--300.

\bibitem[Bartholomew et al.(2011)]{bartholomew2011}
Bartholomew, D. J., Knott, M., \& Moustaki, I. (2011). \emph{Latent variable models and factor analysis: A unified approach} (Vol. 899). John Wiley \& Sons.

\bibitem[Bock \& Aitkin(1981)]{bock1981}
Bock, R. D., \& Aitkin, M. (1981). Marginal maximum likelihood estimation of item parameters: Application of an EM algorithm. \emph{Psychometrika}, 46(4), 443--459.

\bibitem[Boker et al.(2011)]{boker2011}
Boker, S., Neale, M., Maes, H., Wilde, M., Spiegel, M., Brick, T., ... \& Fox, J. (2011). OpenMx: An open source extended structural equation modeling framework. \emph{Psychometrika}, 76(2), 306--317.

\bibitem[Breslow \& Clayton(1993)]{breslow1993}
Breslow, N. E., \& Clayton, D. G. (1993). Approximate inference in generalized linear mixed models. \emph{Journal of the American Statistical Association}, 88(421), 9--25.

\bibitem[Cai(2010a)]{cai2010a}
Cai, L. (2010a). High-dimensional exploratory item factor analysis by a Metropolis-Hastings Robbins-Monro algorithm. \emph{Psychometrika}, 75(1), 33--57.

\bibitem[Cai(2010b)]{cai2010b}
Cai, L. (2010b). Metropolis-Hastings Robbins-Monro algorithm for confirmatory item factor analysis. \emph{Journal of Educational and Behavioral Statistics}, 35(3), 307--335.

\bibitem[Carroll et al.(2006)]{carroll2006}
Carroll, R. J., Ruppert, D., Stefanski, L. A., \& Crainiceanu, C. M. (2006). \emph{Measurement error in nonlinear models: a modern perspective}. Chapman and Hall/CRC.

\bibitem[Chalmers(2012)]{chalmers2012}
Chalmers, R. P. (2012). mirt: A multidimensional item response theory package for the R environment. \emph{Journal of Statistical Software}, 48(6), 1--29.

\bibitem[De Boeck \& Wilson(2004)]{deboeck2004}
De Boeck, P., \& Wilson, M. (Eds.). (2004). \emph{Explanatory item response models: A generalized linear and mixed model approach}. Springer Science \& Business Media.

\bibitem[Murphy(2012)]{murphy2012}
Murphy, K. P. (2012). \emph{Machine learning: A probabilistic perspective}. MIT Press.

\bibitem[Muthén(1984)]{muthen1984}
Muthén, B. (1984). A general structural equation model with dichotomous, ordered categorical, and continuous latent variable indicators. \emph{Psychometrika}, 49(1), 115--132.

\bibitem[Muthén(2002)]{muthen2002}
Muthén, B. O. (2002). Beyond SEM: General latent variable modeling. \emph{Behaviormetrika}, 29(1), 81--117.

\bibitem[Rabe-Hesketh et al.(2004)]{rabe2004}
Rabe-Hesketh, S., Skrondal, A., \& Pickles, A. (2004). Generalized multilevel structural equation modeling. \emph{Psychometrika}, 69(2), 167--190.

\bibitem[Rasmussen \& Williams(2006)]{rasmussen2006}
Rasmussen, C. E., \& Williams, C. K. I. (2006). \emph{Gaussian processes for machine learning}. MIT Press.

\bibitem[Rosseel(2012)]{rosseel2012}
Rosseel, Y. (2012). lavaan: An R package for structural equation modeling. \emph{Journal of Statistical Software}, 48(2), 1--36.

\bibitem[Sammel et al.(1997)]{sammel1997}
Sammel, M. D., Ryan, L. M., \& Legler, J. M. (1997). Latent variable models for mixed discrete and continuous outcomes. \emph{Journal of the Royal Statistical Society: Series B (Statistical Methodology)}, 59(3), 667--678.

\bibitem[Skrondal \& Rabe-Hesketh(2004)]{skrondal2004}
Skrondal, A., \& Rabe-Hesketh, S. (2004). \emph{Generalized latent variable modeling: Multilevel, longitudinal, and structural equation models}. Chapman and Hall/CRC.

\end{thebibliography}
\end{document}